\documentstyle[sprocl]{article}

\def\NPB{{\em Nucl. Phys.} B}

\newcommand{\N}{{\cal N}}
\newcommand{\R}{{\bf R}}

\begin{document}

\title{\vspace{-1in}\parbox{\linewidth}{\small\hfill
CALT-68-2190}
\vspace{0.3in}\\
SUPERSYMMETRIC GAUGE THEORIES\\ AND \\ 
GRAVITATIONAL INSTANTONS}

\author{SERGEY A. CHERKIS}

\address{California Institute of Technology, Pasadena\\ 
CA 91125, USA\\E-mail: cherkis@theory.caltech.edu} 


\maketitle\abstracts{ Various string theory realizations of three-dimensional
gauge theories relate them to gravitational instantons \cite{us}, Nahm equations \cite{us2} and
monopoles \cite{us1}. We use this correspondence to model self-dual gravitational 
instantons of $D_k$-type as moduli spaces of singular monopoles, find their twistor spaces 
and metrics.}

This work provides yet another example of how string theory unites seemingly
distant physical problems. (See references for detailed results.) The central object considered here is supersymmetric gauge theories in three dimensions. In particular, we shall be interested in their
vacuum structure. The other three problems that turn out to be closely related to these gauge theories are:
\begin{itemize}
  \item Nonabelian {\em monopoles} of Prasad and Sommerfield, which are solutions of the Bogomolny equation $$*F=D\Phi$$ (where F is the field-strength of a nonabelian connection $A=A_1 d x^1+A_2 d x^2+A_3 d x^3$ and $\Phi$ is a nonabelian Higgs
field).
  \item An integrable system of {\em equations} named after {\em Nahm} 
\begin{equation}\label{Nahm}
\frac{d T_i}{d s}=\frac{1}{2}\varepsilon_{ijk}[T_j, T_k],
\end{equation}
for $T_i(s)\in u(n).$ These generalize Euler equations for a rotating top.
  \item Solutions of the Euclideanized vacuum Einstein equation called {\em 
self-dual gravitational instantons}, which are four-dimensional manifolds with self-dual curvature tensor
$$R_{\alpha\beta\gamma\delta}=\frac{1}{2}\varepsilon_{\alpha\beta\mu\nu}R^{\mu\nu}_{\ \ \gamma\delta}.$$
\end{itemize}
The latter provide compactifications of string theory and supergravity that preserve supersymmetry and are of importance in euclidean quantum gravity.
The compact examples are delivered by a four-torus and K3. The noncompact
ones are classified according to their asymptotic behavior and topology.
Asymptotically Locally Euclidean (ALE) gravitational instantons asymptotically 
approach $\R^4/\Gamma$ ($\Gamma$ is a finite subgroup of $SU(2)$). These were classified by Kronheimer into two infinite ($A_k$ and $D_k$) series and three exceptional ($E_6$, $E_7$ and $E_8$) cases according to the intersection matrix of their two-cycles. Asymptotically 
Locally Flat (ALF) spaces approach the $\left(\R^3\times S^1\right)/\Gamma$ metric. (To be more precise $S^1$ is Hopf fibered over the two-sphere at infinity
of $\R^3$.)
Sending the radius of the asymptotic $S^1$ to infinity we recover an ALE space
of some type, which will determine the type of the initial ALF space.
For example, the $A_k$ ALF is a (k+1)-centered multi-Taub-NUT space.
Here we shall seek to describe the $D_k$ ALF space.

M theory on an $A_k$ ALF space is known to describe (k+1) D6-branes of type IIA string theory. Probing this background with a D2-brane we obtain an $\N=4$
$U(1)$ gauge theory with (k+1) electron in the D2-brane worldvolume. As the D2-brane corresponds to
an M2-brane in M theory, a vacuum of the above gauge theory corresponds to a position of the M2-brane on the $A_k$ ALF space we started with. Thus the
moduli space of this gauge theory is the $A_k$ ALF. Next, considering M theory
on a $D_k$ ALF one recovers \cite{Sen} k D6-branes parallel to an orientifold $O6^-$. On a D2-brane probe this time we find an $\N=4$ SU(2) gauge theory with k matter multiplets. Its moduli space is the $D_k$ ALF. So far 
we have related {\em gauge theories} and {\em gravitational instantons}.

\begin{figure}[h]
\setlength{\unitlength}{0.9em}
\begin{center}
\begin{picture}(17,13)
\put(3,2){\line(0,1){10}}
\put(8,2){\line(0,1){10}}
\put(3,9){\line(1,0){5}}
\put(3,7){\line(1,0){5}}
\put(5.5,9.7){\makebox(0,0){D3}}
\put(8,11){NS5}
\put(3,11){NS5}

\multiput(8,4)(0,0.8){3}{\line(1,0){6}}
\put(8,10){\line(1,0){6}}
\multiput(11,7)(0,1){3}{\circle*{.1}}
\put(11.5,7.5){k}
\put(11,10.7){\makebox(0,0){D3}}

\end{picture}
\end{center}
\caption{The brane configuration corresponding to $U(2)$ gauge theory with 
k matter multiplets on the internal D3-branes.}
\end{figure}
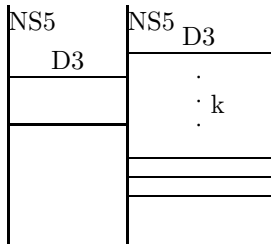
There is another way of realizing these gauge theories. Consider the Chalmers-Hanany-Witten configuration in type IIB string theory (Figure~1).
In the extreme infrared limit the theory in the internal D3-branes will
appear to be three-dimensional with $\N=4$ supersymmetry.
This realizes the gauge theory we are interested in. A vacuum of this theory
describes a particular position of the D3-branes. In the $U(2)$ theory on the NS5-branes the internal D3-branes appear as nonabelian monopoles, while
every external semiinfinite D3-brane appears as a Dirac monopole in the
$U(1)$ of the lower right corner of the $U(2)$. Thus the moduli space of such
{\em monopole configurations} of nonabelian charge two and with k singularities
is also a moduli space of the {\em gauge theory} in question.

Another way of describing the vacua of the three-dimensional theory on the D3-branes is by considering the reduction of the four-dimensional theory on the interval. For this reduction to respect enough supersymmetry the fields (namely the Higgs fields of the theory on the D3-branes) should
depend on the reduced coordinate so that Nahm Equations (\ref{Nahm}) are satisfied. Thus the Coulomb branch of the three-dimensional {\em gauge theory}
is described as a moduli space of solutions to {\em Nahm Equations}.

At this point we have two convenient descriptions of $D_k$ ALF space as a moduli
space of solutions to Nahm equations and as a moduli space of singular 
monopoles. It is the latter description that we shall make use of here. 
Regular monopoles can be described \cite{AH} by considering a scattering 
problem $\left(\vec{u}\cdot(\vec{\partial}+\vec{A})-i\Phi\right)s=0$\  on every
line $\gamma$ in the three-dimensional space directed along $\vec{u}$. The space
of all lines is a tangent bundle to a sphere ${\bf T}={\bf TP}^1$. Let $(\zeta,\eta)$ be standard coordinates on ${\bf T}$, such that $\zeta$ is a coordinate on the sphere and $\eta$ on the tangent space. Then the set of lines
on which the scattering problem has a bound state forms a curve $S\in{\bf T}$.
$S$ is called a spectral curve and it encodes the monopole data we started with.

In case of singular monopoles some of the lines $\gamma\in S$ will pass through 
the singular points. These lines define two sets of points $Q$ and $P$ in $S$, such that $Q$ and $P$ are conjugate to each other with respect to the change of orientation of the lines. Analysis of this situation \cite{us} shows, that in addition to 
the spectral curve, we have to consider two sections $\rho$ and $\xi$ of the line
bundles over $S$ with transition functions $e^{\mu\eta/\zeta}$ and $e^{-\mu\eta/\zeta}$
correspondingly. Also $\rho$ vanishes at the points of $Q$ and $\xi$ at those of $P$.

Since we are interested in the case of two monopoles the spectral curve is given
by $\eta^2+\eta_2(\zeta)=0$ where $\eta_2(\zeta)=z+v\zeta+w\zeta^2-\bar{v}\zeta^3+\bar{z}\zeta^4$. z,v and w are the moduli. z and v are complex and w is real. The sections $\rho$ and $\xi$
satisfy $\rho \xi=\prod_{i=1}^k\left(\eta-P_i(\zeta)\right)$, where $P_i$ are 
quadratic in $\zeta$ with coefficients given by the coordinates of the singularities. The above equations provide the description of the twistor space 
of the singular monopole moduli space. Knowing the twistor space one can use the 
generalized Legendre transform techniques \cite{LR} to find the auxiliary
function $F$ of the moduli
\begin{eqnarray} \nonumber
F(z,\bar{z},v,\bar{v},w)&=&\frac{1}{2\pi i}\oint_0 d\zeta \frac{2 \mu \eta_2}{\zeta^3}+2\oint_{\omega_r}d\zeta
\frac{\sqrt{-\eta_2}}{\zeta^2}-\\
&&\sum_a\frac{1}{2\pi i}\oint_{\! C_a}\ \frac{d\zeta}{\zeta^2}
(\sqrt{-\eta_2}-z_a(\zeta))\log (\sqrt{-\eta_2}-z_a(\zeta)).
\end{eqnarray}
Imposing the consistency constraint $\partial F/\partial w=0$ expresses $w$ as
a function of $z$ and $v$. Then the Legendre transform of $F$
\begin{equation}
K(z,\bar{z},u,\bar{u})=F(z,\bar{z},v,\bar{v})-uv-\bar{u}\bar{v},
\end{equation}
with $\partial F/\partial v=u$ and $\partial F/\partial \bar{v}=\bar{u}$,
gives the K\"ahler potential for the $D_k$ ALF metric. This agrees with the conjecture of Chalmers \cite{Chalmers}.

\section*{Acknowledgments}
The results presented here are obtained in collaboration with Anton Kapustin.
This work is partially supported by  DOE grant DE-FG03-92-ER40701.

\end{document}